# Optimizing the Cost for Resource Subscription Policy in IaaS Cloud

Ms.M.Uthaya Banu[#1], Mr.K.Saravanan[*2]

[#] *Student,* [*] *Assistant Professor*
*Department of Computer Science and Engineering*
*Regional Centre of Anna University, Tirunelveli (T.N) India*

*Abstract*—Cloud computing allow the users to efficiently and dynamically provision computing resource to meet their IT needs. Cloud Provider offers two subscription plan to the customer namely reservation and on-demand. The reservation plan is typically cheaper than on-demand plan. If the actual computing demand is known in advance reserving the resource would be straightforward. The challenge is how to make properly resource provisioning and how the customers efficiently purchase the provisioning options under reservation and on-demand. To address this issue, two-phase algorithm are proposed to minimize service provision cost in both reservation and on-demand plan. To reserve the correct and optimal amount of resources during reservation, proposed a mathematical formulae in the first phase. To predict resource demand, use kalman filter in the second phase. The evaluation result shows that the two-phase algorithm can significantly reduce the provision cost and the prediction is of reasonable accuracy.

*Keywords*— Pricing and Resource Allocation, Prediction.

## I. INTRODUCTION

### A. BACKGROUND

Cloud computing is a technology that uses the internet and central remote servers to maintain data and applications. Cloud computing allows consumers and business to use applications without installation and access their personal files at any computer with internet access. It allows for much more efficient computing by centralizing storage, memory, processing and bandwidth. According to NIST states "cloud computing is a model for enabling convenient, on-demand network access to a shared pool of configurable computing resources (e.g., networks, servers, storage, applications, and services) that can be rapidly provisioned and released with minimal management effort or service provider interaction". One fundamental advantage of the cloud paradigm is computation outsourcing, where their resource-constraint devices no longer limit the computational power of cloud customers. By outsourcing the workloads into the cloud, customers could enjoy the literally unlimited computing resources in a pay-per-use manner without committing any large capital outlays in the purchase of both hardware and software and/or the operational overhead therein. Each provider serves a specific function, giving users more or less control over their cloud depending on the type. When consumers choose a provider, they compare their needs to the cloud services available. Cloud consumer needs will vary depending on how they intend to use the space and resources associated with the cloud. Keep in mind that cloud provider will be pay-as-you-go, meaning that if technological needs change at any point consumer can purchase more storage space (or less for that matter) from cloud provider. The cloud computing model is comprised of a front end and a back end. These two elements are connected through a network. The front end is the vehicle by which the user interacts with the system and the back end is the cloud itself. The front end is composed of a client computer, or the computer network of an enterprise, and the applications used to access the cloud. The back end provides the applications, computers, servers, and data storage that creates the cloud of services. Cloud computing describes a type of outsourcing of computer services, similar to the way in which the supply of electricity is outsourced. Users can simply use it. They do not need to worry where the electricity is from, how it is produced, or transported. In cloud, services allowing users to easily access resources anywhere anytime. Users can pay for a service and access the resources made available during their subscriptions until the subscribed periods expire. Users are then forced to demand such resources if they want to access them also after the subscribed periods. We mainly focused on the service provision issues on IaaS, which abstracts hardware resources into pool of computing resources and virtualization infrastructure. IaaS providers build flexible cloud solutions according to the hardware requirements of customers; furthermore it let customers run operating systems and software applications on virtual machine (VMs).Customers merely pay for the resources that are actually used. To host web application services, service operators would apply resource subscription plans to dynamically adjust service capacity to satisfy a time-varying demand. While subscribing IaaS resources, the web service operators aimed to provide a certain level Agreement (SLA) with their clients, e.g.,a guarantee on request response time. The resource provisioning of IaaS allows consumers to elastically increase or decrease the system capacity by changing configurations of computing resources. Moreover, cloud providers have multiple usage-based pricing models based on different VM configurations, such as different CPU cores, memory size, and rental costs.

Cloud providers generally offer at least two subscription plans to their customers.(i.e., reservation and on-demand plans)to their customers. The on-demand plan is typically more expensive than the reservation plan because the former allows VMs to be dynamically acquired at anytime without a commitment and charged on a pay-per-use-basis. On the other hand, with the reservation plan, users need to pay an upfront fee for the contract. Then, the reserved VMs can be utilized at a cheaper usage cost during the time of contract. In this way, customers achieve significant cost savings. However, there are some unpredictable situations, such as uncertain demand, incurring over- and under- provisioning problems. For example, the time-varying workload fluctuation increases the difficulty of demand estimation. Owing to the error-prone demand estimation and complex combination of cloud resources, customers usually make inappropriate subscription plans. Clearly, the resource subscription problem can be divided into two sub-problems: how many long-term resources to be reserved and how many on-demand resources to be acquired. If the long term reserved resources are more than the actual demand, it causes waste of the upfront fee. On the other hand, if the reserved resources are less than the actual demand, additional resources need to be subscribed on-demand, which are more expensive than usage cost of long term reserved resources.To alleviate this problem, a promising mechanism is to prepare extra resources ahead of time by predicting traffic demand. Furthermore, different ways of increasing resources also need to be taken into account, such as VM migration and replication.

*B. MOTIVATION AND OBJECTIVE*

IaaS is capable of dynamically providing virtual infrastructure according to the demand of users and offering flexible provisioning plans In IaaS cloud environment, a variety of computing resources can be combined to form different types of VM, each with a different combination of capacities of different resources. There are three different rental costs, including an upfront fee for long term reservation, a usage charge of reserved resources, and an on-demand cost on dynamically allocated resources

Our objective is to minimize the operational cost by virtue of optimal resource reservation and predictive adjustment of resource usage. The following techniques were used to achieve the objective:

1) For Long term resource reservation, aimed to find the amount of resources to be leased such that the operational cost could be minimized, assuming that insufficient resources at any time instance could be dynamically and instantaneously allocated on demand. The resource reservation plan included the lease period, types of VM and their quantity to be reserved.

2) For on-demand resource allocation, adopted the kalman Filter to predict workload demand; the VM configuration problem was formulated. The VM configuration problem took into account of VM launch or shutdown to reflect the change of workload demand.

## II. RELATED WORK

This research work explores the issues to application providers on how to effectively provision or subscribe VM resources from an IaaS provider, areas related to other work include the following:1)Pricing model of cloud resources;2)Resources Provisioning for cloud computing; and 3)Resource Demand Prediction.

In[1], Ren-Hung Hwang et al. proposed two subscription plan namely long-term reservation and on-demand subscription.To make properly resource provisioning and minimize service provision cost, two phase algorithm were used.

In [2], Roussopoulos et al. proposed a microeconomic inspired approach to determine the number of VMs to be allocated to each user by their financial capacity, then maximizing per-user's profit and effectiveness of sharing.

In earlier work such as [3], applied a queuing model to analyze the response time distribution for two classes of job; moreover, a heuristic algorithm was developed to obtain the smallest number of servers without violating the SLAs.Howerver, in a real world situation, most of IaaS providers offer several pricing options: reservation, on-demand, and spot. Reservation option is suitable for long term provisioning. A customer signs a long term lease contract with the IaaS provider to reserve a fixed amount of resources. Usually, it includes an upfront fee for signing the contract and a usage fee per instance and unit of time for actual use of the resources. On-demand option is suitable for dynamic provisioning. A customer can ask for resources on a pay-peruse- basis at any time. The Amazon EC2 also provides the spot option, which allows a customer to submit a bid price to compete with remained resources.

Calheiros et al. [4] analyzed a provisioning technique that automatically adapts to workload changes related to applications for facilitating the adaptive management of system and offering end-users guaranteed Quality of Services (QoS) in large, autonomous, and highly dynamic environments. They model the behavior and performance of applications and Cloud-based IT resources to adaptively serve end-user requests. To improve the efficiency of the system, they use analytical performance (queueing network system model) and workload information to supply intelligent input about system requirements to an application provisioner with limited information about the physical infrastructure.

Mao et al. [5] proposed a cloud auto-scaling mechanism to automatically scale computing instances based on workload information and performance desire. The mechanism schedules VM instance startup and shut-down activities. It

ISSN: 2231-5381           http://www.ijettjournal.org                Page 297

enables cloud applications to finish submitted jobs within the deadline by controlling underlying instance numbers and reduces user cost by choosing appropriate instance types. They have implemented mechanism in Windows Azure platform, and evaluated it using both simulations and a real scientific cloud application.

Chaisiri et al. [6] studied the the optimal long term reservation plan and on-demand plan. They formulated the optimal long term reservation plan problem as a stochastic programming model by assuming that the distribution of workload demand and price model was known in prior. However, the computation complexity is too high and the assumption is not practical.

In [7], two provisioning algorithm for long-term and short-term planning were proposed. The long-term plan was determined by subscribing reserved instances for the long-term usage. Inside the long-term plan, multiple short-term plans were triggered to provide enough numbers of spot instances for capacity supporting. Even though they developed the short-term plan, they overlooked the time delay to launch a new VM.

Mark et al. [8] adopted a demand forecaster to predict the future workload such that occurrences of over provisioning could be reduced. Nevertheless, their prediction mechanism was only used in the reservation phase and the delay of dynamic resource provision was neglected in the on-demand phase.

### III. PROPOSED SYSTEM

For long-term provisioning, we aim to determine the optimal number of VMs needs to be reserved. For short-term provisioning, we aim to determine how to configure VMs to provide sufficient service capacity for time varying workload.

We assume that an IaaS provider offers multiple VM types. Each type features different hardware specifications. Besides, there are three different rental costs, including an upfront fee for long term reservation, a usage charge of reserved resources, and an on-demand cost on dynamically allocated resources. These costs are normalized to per short-term time interval hereafter for ease of cost calculation.

Let $V = \{V1, V2, …, VM\}$ denote the set of VM types and M be the total number of VM types supported by the IaaS provider. Each VM type has its own hardware specification and service capacity. Let $C_i$ denote the capacity of $V_i$ which corresponds to the maximum number of concurrent users or the service request rate that can be supported by an instance of $V_i$ without violating the quality of service guarantee.

### A. SYSTEM ARCHITECTURE

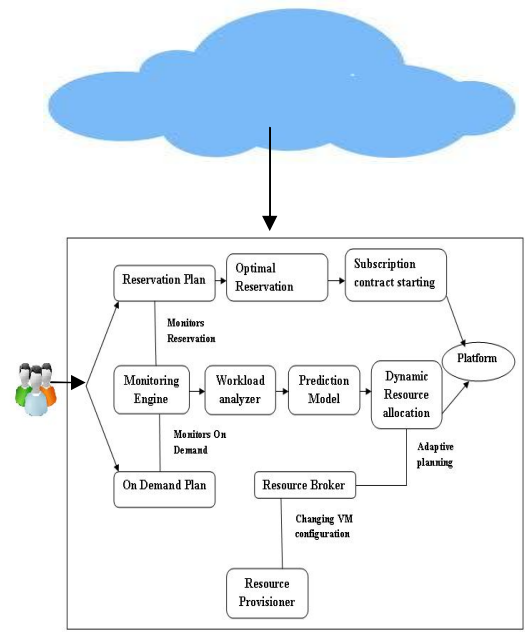

Fig. 1 Operational Overview of System Architecture

The overview of system architecture is shown which consists of two roles: Service Provider and IaaS Provider. The Resource Provisioner component of IaaS Provider takes the responsibility for adjusting the deployment of VM repository based on the Service Provider's request. The Service Provider consists of several key components which include Monitoring Engine, Workload Analyzer, Prediction Model, Elasticity planner and Resource Broker. The IaaS providers consist of Resource Provisioner.

*1) Monitoring Engine:* The Service Provider which traces the number of simultaneous online user and resource utilization. The demand fluctuates over the monitoring time.Fluctutations of demand would cause an under provisioning if we only rely on long term reserved resources and for over provision case, that is when workload demand is less than the reserved resource, attention is required for the usage charge of launching a reserved VM.

*2) Workload Analyzer*: The Provider which generates the analysis of workload based on user request, resource utilization and their duration. Analysis for each activity of all individual users for several days. Work measurement entrusted to each users with time intervals, i.e., daily, weekly, monthly.

*3) Prediction Model:* This model which makes use of Kalman filter according to the analysis of workload to provide demand prediction.

*4) Resource Broker:* This model which performs adaptive planning and delivers resource subscription to the IaaS Provider.

IV. METHODOLOGY

The optimization process has two phases, and their main functionalities are long-term resource reservation optimization and effective short-term resource allocation. In this section, we describe the proposed two-phase planning algorithms.

*A. Long term Resource Reservation*

In the long-term resource reservation,

1. Given set of demands we have to calculate the provisioning cost which includes the upfront fee, the usage charge for launching reserved VMs when demand is less than reservation capacity, the usage charge of launching all the reserved capacity when the demand exceeds the reserved capacity and the cost for on-demand allocated VMs to serve the exceeded demand.

2. The objective is to minimize the provisioning cost and to derive the optimal amount of long-term reserved resource with a model where the demand is a discrete random variable and only one single type of VM is considered.

3. Assume that $r^*$ is the optimal number of VMs to be reserved for long term planning and calculate an upper and lower bound of the optimal number of reserved VMs.

4. To show how to use the result of single type VM solution for the original problem with multiple VM types.

(i) Selecting the VM that has the best capacity/price (CP) ratio

(ii) Under constraint (i), the workload demand is transformed to demand of the number of best CP ratio VMs

(iii) Based on upper and lower bound we will obtain the optimal reservation

(iv) We could do a search for the best combination of multiple VM types with capacity falls between the capacity of $(r^*-1).C_{bestcp}$ and $r^*.C_{bestcp}$.

$C_{bestcp}$ is the capacity of the VM with the best CP ratio.

(v) Each value in the range is regarded as the reserved demand of the Integer Linear Programming formulation.

Minimize
$$\sum_i^M n_i * p_i^R \quad (1)$$
Subject to
$$\sum_i^M n_i * C_i \geq \text{Reserved Demand} \quad (2)$$
$$n_i \in N_o \quad i \in M \quad (3)$$

(1)→to minimize the upfront cost of reserved VMs

$n_i$→number of type i VM that is subscribed in the long term lease contract.

*B. On Demand Resource Allocation*

A straight forward way to configure VMs for next short-term planning interval is based on the measured demand of current $I_0$ configuring VMs based on some prediction mechanism will significantly reduce the operational cost. Our prediction mechanism is based on kalman filter because it has low computation complexity.

**SHORT-TERM PLANNING ALGORITHM (SPA)**

In the following, describe the proposed short-term planning algorithm (SPA). Depending on the values of *rp*, *rc*, and *rr*, the SPA classifies the resource planning scenarios into three cases which are illustrated as follows:

**Input:**

$r_m, r_p, r_c, r_r, I_c, I_r$

$r_m$->VM Capacity requirement for current demand

$r_p$->Predictive VM Capacity

$r_c$->Current launched VM Capacity

$r_r$->Overall VM Capacity of reserved resources

$I_c$→Current launched VM Configuration

$I_r$->VM Configuration in reservation contract

**Output:**

The updated $I_c$, which is used for adaptive planning

**Initialization:**

$I_O := \{0\}$ //VM Configuration subscribed via on-demand plan is empty

$I_O$ -> The VM configuration subscribed via on-demand plan

**Procedure:**

1 if $r_r < r_p$ then

2 $I_o$ ←ILP1_OnDemand ($r_p - r_r, \Delta$)

3 $I_c = I_r + I_o$

4 else if $r_c < r_p$ then //launching more reserved VMs is required

5 $I_c$ ←ILP2_AdjustVMConfiguration ($I_r, \Delta, I_c, r_p$)

6 else if $r_c > r_p$ then

7 $I_c$ ←ShutDownSpareVMs ($I_c, r_p$)

8 end if

9 UpdatePredicationModel ($r_m$)

10 return $I_c$

End procedure

Depending on the values of *rp*, *rc*, and *rr*, the SPA classifies the resource planning scenarios into three cases which are illustrated as follows:

**Scenario 1 (lines 1-3): On Demand Resource**

The predicted demand (*rp*) exceeds the capacity of all reserved VMs (*rr*), thus the Resource Broker must operate the on-demand option to subscribe more VMs.

**Scenario 2(lines 4-5): Adjust VM configuration**
The predicted demand can be served by reserved VMs, but it exceeds the capacity of current VM configuration (*rc*). Therefore, reconfiguring launched VMs from the reserved VM pool (*Ir*) is necessary.

**Scenario 3 (lines 6-7): Shutdown spare VMs**
The predicted demand is less than the currently configured VM capacity. Therefore, the corresponding action is to shut down some launched VMs, which had nearly a full hour of operation first, until the provisioning capacity is just above the predicted demand.

## V. RESULTS

The Results for long-term reservation and on-demand cost is shown in following screen shots.

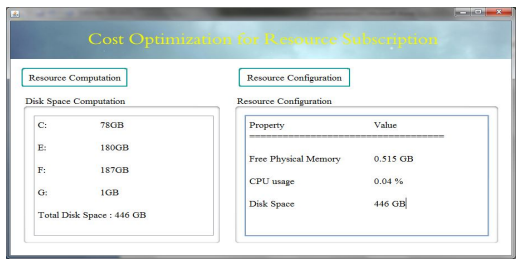

Fig. 2 Resource Computation and Resource Configuration of Virtual Server1

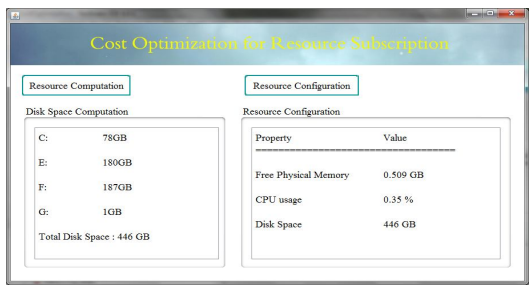

Fig. 3 Resource Computation and Resource Configuration of Virtual Server2

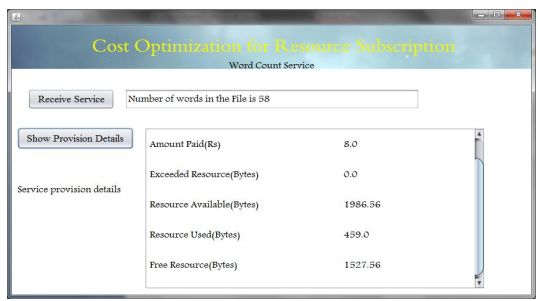

Fig. 4 Service Provision Details

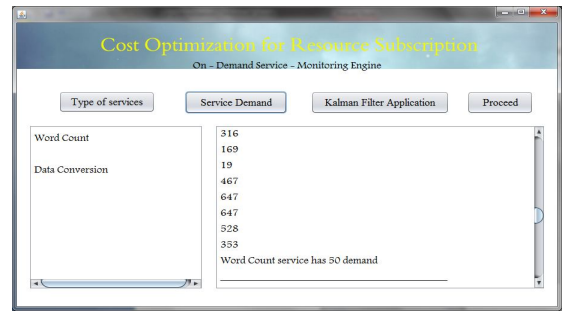

Fig. 5 Monitoring Engine

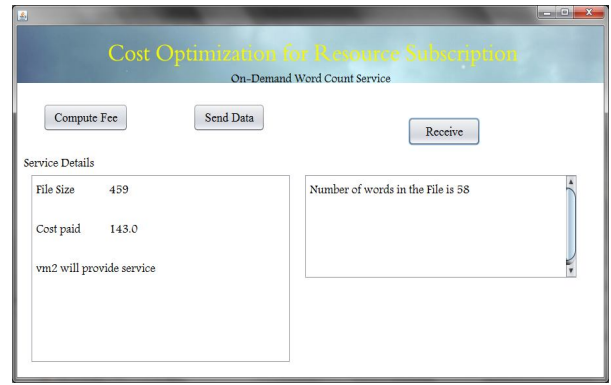

Fig. 5 On Demand Service

## VI. EVALUATION

In this section, the experimental evaluation of the proposed algorithm is presented. We first present the parameter settings of a cloud computing environment used in this performance evaluation. We adopt the pricing models set by Amazon EC2. Since we re-configure VMs every short-term planning interval, we present the normalized upfront and usage costs per short-term planning interval. In order to evaluate the performance of Resource Subscription policy in cloud, the following schemes are used:1)Long-Term Cost Computation, and 2)VM Allocation.

### A. Long-Term Cost Computation

In this scheme Customers are requested to use their service based on duration and the resource will be allotted for that particular duration and the upfront cost is collected from the customer in order to get their service. Each user can upload their file and get their requested Service and the service provisioning cost will be collected from the customer. When the resource exceeds while using the service during their duration, normally resource will be allotted from the on-demand. But in proposed algorithm, based on the user needs resource will be extended normally and the cost will be computed for that particular resource usage. This is generally cheaper than on-demand Resources. The following graph which represents the comparison of cost based on extended resources and based on on-demand resources.

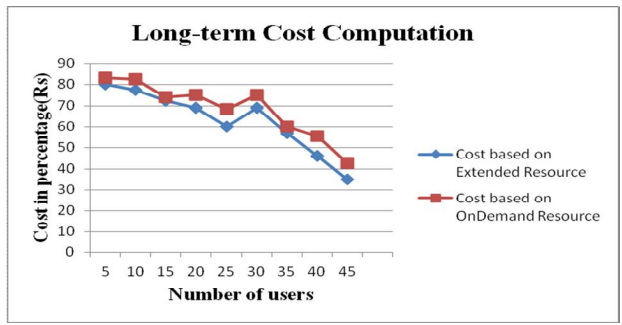

Fig. 6 Long Term Cost Computation

*B. Comparing VM Allocation*

In this scheme Resource capacity is calculated first and then resources are configured automatically based on CPU usage and memory capacity. Here we considered two VMs and we calculated service for different users and also the allocation of VM for these different users. The following graph shows the comparison of VM allocation.

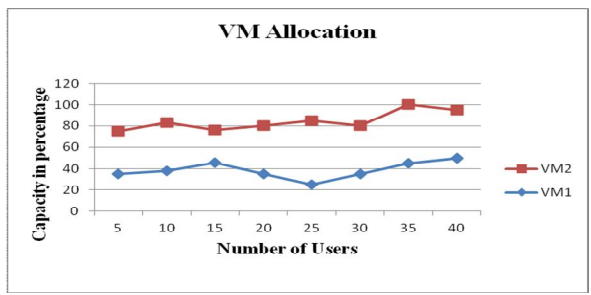

Fig. 7 VM Allocation

## VII. CONCLUSION

IaaS infrastructure becomes a popular platform for application providers to deploy their applications. However, IaaS providers offer many types of VM configuration and price them differently. Furthermore, they also offer several pricing models. It raises an interesting issue to application providers on how to effectively provision or subscribe VM resources from an IaaS provider. In this paper, we formulated the resource provisioning problem as a two phase resource planning problem. In the first phase, focused on determining the optimal long term resource provisioning.In the second phase, proposed a Kalman filter prediction model for predicting resource demand. Several issues had also been considered in our work, including impact of latency of VM re-configuration, and minimum rental time constraint for launching a VM. Our numerical results showed that the proposed long term resource planning algorithm was able to yield near optimal operational cost. The results also showed that the proposed on-demand planning algorithm significantly reduced the operational cost. In future, we plan to evaluate our solutions with larger resource demand from some real web application system and were able to cope with the latency of VM reconfiguration.